\documentclass[11pt,a4paper]{article}  
\usepackage{graphicx}
\usepackage{epstopdf}
\usepackage{amssymb}
\usepackage{stackengine}
\usepackage{mathtools}

\usepackage{slashed}

\def\manifold{H_d}

\def\H{{\cal H}}

\def\Lobad{{H_d}}
\def\n{\kappa}
\def\ho{\H_\omega}

\newcommand{\sd}{{ S}_{d-1}}
\def\formdif{{\alpha}}
\def\wh{\widehat}
\def\wt{\widetilde}

\def\ro2{||^2_{\ho}}

\def\beq{\begin{equation}}
\def\endq{\end{equation}}
\def\beqa{\begin{eqnarray}}
\def\endqa{\end{eqnarray}}
\def\TT{{\cal T}}
\def\bC{{\bf C}}
\def\bR{{\bf R}}

\def\M{{\cal M}}

\def\W{{\cal W}}

\def\bea{\begin{eqnarray}}
\def\eea{\end{eqnarray}}

\def\beas{\begin{eqnarray*}}
\def\eeas{\end{eqnarray*}}

\def\P{ P}
\def\T{{\cal T}}

\newcommand{\be}{\begin{equation}}
\newcommand{\ee}{\end{equation}}
\newcommand{\ba}{\begin{eqnarray}}
\newcommand{\ea}{\end{eqnarray}}

\renewcommand{\Bbb}{\bf}
\renewcommand{\Re}{\mathrm {Re} \,}
\renewcommand{\Im}{\mathrm {Im} \,}

\newtheorem{property}{Property}

\newtheorem{theorem}{Theorem}
\newtheorem{Remark}{Remark}
\newtheorem{Corollary}{Corollary}

\title{Plane waves, harmonic analysis in de Sitter and anti de Sitter Quantum Field Theory and the spectral condition}

\author{Ugo Moschella $^{1,2,3}$ \\
$^{1}$ Universit\`a dell'Insubria,  Como, Italia \\
$^{2}$  CERN, Theory Department, Geneva, Switzerland.\\
$^{3}$ INFN Sez. di Milano, Via Celoria 16, 20133 Milano, Italia}

\begin{document}


\maketitle
\abstract{{}\footnote{THE FRIEDMANN COSMOLOGY: A CENTURY LATER. Invited paper} 
We review the role of the spectral condition as a characteristic feature unifying Minkowski, de Sitter and anti de Sitter Quantum Field Theory. In this context, we highlight the role of an important class of plane waves which are either de Sitter or anti de Sitter covariant and are compatible with the relevant analyticity domains linked to the spectral condition(s). We show again how to expand the two-point functions and propagators in terms of them and some of the advantages of doing so rather than using special coordinate systems and separated variables.}

\section{The birth of the de Sitter model}

After writing, in December 1915,  his equations for the geometry of spacetime, Einstein turned his attention to cosmology and tried to apply them to the entire universe, creating an entirely new science:  modern scientific cosmology, whose founding idea is that a global exact solution of Einstein's equations corresponds somehow to a model for the universe. 

Einstein's concern was at first epistemological:
the metric structure of the universe must be entirely determined by the material content -- this is more or less the so-called Mach principle. 
But general relativity still keeps a remnant of absolute space in the boundary conditions that must be specified at spacelike infinity to determine the spacetime geometry. To solve this problem, or rather, to dispose of it,  Einstein's ``crazy idea''  was: to let the universe be spherical, let it have spherical spatial sections.

A curved sphere should be imagined as a three-dimensional spherical hypersurface embedded in a Euclidean space of dimension four:
\begin{equation}
S_3 = \{x_1^2+x_2^2+x^2_3+x_4^2= r^2\}. \label{sphere}
\end{equation}
It obviously has no center, or rather it has its center everywhere\footnote{{\em Sphaera infinita cuius centrum est ubique, circumferentia tamen nullibi} : this is the second definition of God that can be read in the Liber XXIV philosophorum, an anonymous medieval treatise attributed to Hermes Trismegistus. Nicolas de Cues applies this definition to the universe: The world machine has, so to speak, its center everywhere and its circumference nowhere (La Docte Ignorance, 1440).  Giordano Bruno at many different places later takes up the definition. Einstein's novelty is that his sphere is not infinite, but finite and curved.} and any point is equivalent to any other point. It has no boundary either; and therefore: no boundary, no conditions on the boundary. 

There was also  a second guiding principle in Einstein's cosmological research:  the universe had to be static, its geometry should not change as time goes by.  In 1917 the visible universe still coincided with the Milky Way, the nebulae enigma had not yet been solved and the hypothesis of a static universe was perfectly reasonable. 
But his General Theory of Relativity of 1915 does not allow for spherical static solutions. 

Here came idea that would be remembered as his {\em biggest blunder}: to add to his equations a constant term  $\Lambda$  that acts repulsively and counteract the gravitational attraction.  That was 
\begin{quote}
{\em an extension of the equations which is not justified by our real knowledge of gravitation} $[\ldots] $ {\em this term is necessary only for the purpose of making possible a quasi-static distribution of matter as required by the low speed of stars} \cite{ein}. 
\end{quote}
This commentary indicates that already in his 1917 paper Einstein was aware of the fact that his original equations of 1915 implied a dynamical universe but he had set aside this possibility; he kept adding $\Lambda$ and found a perfectly Machian static spherical solution, his static model of 1917.

Shortly after the publication of Einstein's paper, de Sitter published a second solution of the new cosmological equations of 1917: an otherwise empty universe  made only by the cosmological constant. 
The astronomer  found his model elaborating on the the boundary condition problem: according to him, Eintsein's solution still retained a trace of absolute space; a four dimensional (complex) sphere could solve the problem in a more  convincingly covariant way. 
As for the sphere (\ref{sphere}), also the de Sitter model can be better visualised as an embedded surface:  it is the four-dimensional one-sheeted hyperboloid embedded in a five-dimensional Minkowski spacetime $M_5$
\begin{equation}
dS_4 = \{x_0^2-x_1^2-x_2^2-x^2_3-x_4^2= -R^2\}. \label{dsman}
\end{equation}

Einstein was very unhappy with the new solution but all his attempts to demonstrate that de Sitter's calculations were faulty consistently failed. 
Einstein finally surrendered: the de Sitter universe was indeed a regular solution of his cosmological equations  without matter but, he said, it was nevertheless  without physical interest because not globally static. Einstein was rejecting the possibility of a dynamical universe, other scientists simply did not know: until the early 1930s the fundamental articles published in  1922  \cite{fri1}  and 1924  \cite{fri2} by Friedmann, who made use of the original equations of general relativity  to describe expanding universes, were substantially ignored. 
Lema\^itre's independent work of 1927 \cite{lem},  based on the cosmological equations of 1917, was ignored too.

On a trip to Pasadena Einstein learned of Hubble's latest observations and was persuaded of the advantages of dynamic models to describe the universe. In two articles published shortly afterwards, Einstein asserted that the original reasons for introducing the cosmological constant no longer existed. Farewell to the cosmological constant. 

In 1947,  Einstein's wrote to Lema\^itre:

\begin{quote} {\em The introduction of such a constant implies a considerable renunciation of the logical simplicity of the theory... Since I introduced this term, I had always a bad conscience $\ldots$  I am unable to believe that such an ugly thing should be realized in nature.} 
\end{quote}
 Lema\^itre's answer of 1949  sounds like a prophecy: 
  \begin{quote}{\em The history of science provides many instances of discoveries which have been made for reasons which are no longer considered satisfactory. It may be that the discovery of the cosmological constant is such a case.} 
  \end{quote}
  In fact, Einstein himself had been prophetic in 1917, in a letter to de Sitter: 
 \begin{quote}
 {\em In any case, one thing is clear. The theory of general relativity allows adding the term $\Lambda$ in the equations. One day, our real knowledge of the composition of the sky of fixed stars, the apparent motions of the fixed stars and the position of spectral lines as a function of distance, will probably be sufficient to decide empirically whether or not $\Lambda$ is equal to zero. Conviction is a good motive, but a bad judge.} \end{quote}

 In 1997, exactly eighty years after its discovery, the cosmological constant was observed \cite{riess,perl}; or, maybe, it was something similar that we now call ``dark energy''. These observations have upturned consolidated and rooted ideas, indicating that the gravitational effect of the greatest part of the energy of the universe consists in producing an accelerated expansion, as in the case of Einstein's cosmological constant. 
Nowadays almost every physicist believes that the dark component  constitutes about seventy percent of the energy  of the universe and that its proportion, according to the standard cosmological  $\Lambda$CDM (cold dark matter) model,  is destined to increase. In the end, only the cosmological constant will remain, and the universe will become a perfect de Sitter spacetime. 

The de Sitter geometry seems therefore to assume the role of reference geometry of the universe. In other words  it is de Sitter's, and not Minkowski's, the geometry of spacetime when the latter is deprived of its content of matter and radiation.

Beyond the acceleration of the universe at late times, also the idea of inflation  consists in  a phase of accelerated quasi-exponential expansion, which is approximately described by de Sitter's geometry in  the primordial universe. A theoretical understanding of the structure of the universe which is observable today is based on quantum field theory on the de Sitter spacetime:
quantum  fluctuations of the vacuum at the epoch of inflation are thought to be responsible for the primordial density inhomogeneities which are at the origin of the structures that exist in the universe today. 

Actually, once one admits that a cosmological constant  may exist, it might also be negative, isn't it?
The model  of universe with a negative cosmological constant and nothing else is termed anti-de Sitter.
It is a curious coincidence that in the very same year 1997 also the negative cosmological constant took center stage in theoretical physics \cite{malda} with the formulation of the by-now  famous  AdS/CFT (Anti-de Sitter/Conformal Field Theory) correspondence, a conjectured duality between two different physical theories.  1997: the year of the two cosmological constants!

\section{Quantum field theory: the spectral condition}

The de Sitter and the anti de Sitter spacetimes have thus great importance in contemporary theoretical physics and cosmology and both dS and AdS Quantum Field Theory (QFT) also play a major role. 
The dS and AdS manifolds  share the properties of having constant curvature and being maximally symmetric manifolds. Actually, in the general $d$-dimensional case, they are just different real submanifolds of one and the same complex manifold: the complex $d$-dimensional sphere 
\begin{equation}
S_d^{(c)} = \{z\in {\bf C}^{d+1}; \ \ z_0^2+z_1^2+\ldots + z_d^2= R^2\}.
\end{equation}
Otherwise, their geometries are radically different from each other: in particular, the (real)  de Sitter manifold  has no global timelike Killing vector field while  the (real) anti de Sitter manifold is not globally hyperbolic and has closed timelike curves. One can get rid of those closed curved by moving  to the universal covering of the real AdS manifold (even though this move might be just an illusion...);  but the universal covering remains not globally hyperbolic,

Global hyperbolicity is a basic property of quantum field theory on curved spacetimes as is usually formulated. Its absence renders AdS QFT a little more demanding from a technical viewpoint but, as we will see, this is not a major difficulty, since there exists in AdS the possibility of identifying a global energy operator.
It is precisely the lack of a global energy operator, which is consequence of the absence of a global timelike Killing vector field, which renders dS QFT actually  more difficult.

There is however a unifying  characteristics that makes dS and AdS QFT's similar to each other and similar also to the standard zero temperature Minkowski QFT:  this is the analyticity of the correlation functions in suitable domains of the respective complexified manifolds. This unifying viewpoint is discussed in the following sections. 

Here, to prepare the ground,  we start by recalling that the fundamental theorem of Stone and Von Neumann, which states the uniqueness of the Hilbert space representation of the Canonical Commutation Relations (CCR's), fails for infinite quantum systems: the distinction between observables and states, which is of no consequence for finitely many degrees of freedom, now becomes crucial and there exist infinitely many Hilbert space realizations of the same  algebra of the observables. In other words,  knowing the Lagrangian of a quantum field theory is not enough; the Lagrangian just provides the commutation rules but  there are infinitely many inequivalent solutions of the field equations sharing the same commutation rules; one needs to specify some extra information to find the physically relevant ones. Only after this step has been taken, transition amplitudes may be computed and comparison with the outcomes of the experiments may be done. 
\section{States and two-point functions}
This non uniqueness is true already at the level of free fields. What is unique is the commutator: on a globally hyperbolic manifold $({\cal M},g)$  the Klein-Gordon Lagrangian uniquely selects the (covariant) commutator 
 $C(x_1,x_2)$ which is an {\em antisymmetric} bi-distribution 
 solving the Klein-Gordon equation  in each variable
 \begin{eqnarray} 
(\Box_{x_1} +m^2)C(x_1,x_2)= (\Box_{x_2}+m^2)C(x_1,x_2)=0
\end{eqnarray}
with precise initial condition given by the equal time canonical commutation relations; the equal time CCR   
in turn imply that $C(x_1,x_2) = 0$  for  any two events   $x_1,x_2$ of $ {\M}$ which are spacelike separated  w.r.t. notion of locality inherent to ${\cal M}$.


 \label{scheme}
 For free fields the smeared commutator is a multiple of the identity element of the field algebra (a {\em c-number}):  given two test functions $ f,g $ belonging to a suitable test function space $ {\cal T} ({\cal M}) $
\begin{equation}[\phi(f), \phi(g)] = C(f,g)\,  = \int _{{\M}\times {\M}} C(x_1,x_2)f(x_1) g(x_2) \,   \sqrt{-g(x_1)}dx_1\, \sqrt{-g(x_2)}dx_2.
\label{comm}\end{equation}
A quantization is accomplished when the 
the commutation relations (\ref{comm}) are represented by an operator-valued distribution in a
Hilbert space $\cal H$:  one should determine a linear map 
\begin{equation}
\phi (f) \longrightarrow \widehat \phi (f)\in Op({\cal H}) 
\end{equation}
preserving the algebraic structures and such that 
\begin{equation}
 [\widehat \phi (f),\widehat \phi (g)] = C(f,g)\, {\mathbf 1} \label{poi1}.
\end{equation}
As we said,  the Stone-Von Neumann theorem fails and there are  uncountably many solutions to this problem. 
How can we construct (at least some of) them?

A possible solution is completely encoded in the knowledge of a  two-point function  i.e. 
a two-point 
distribution  $\W \in {\cal T}' ({\cal M} \times {\cal M})$
that solves  the Klein-Gordon equation  in each variable 
\begin{equation}
\left(\square_{x_1}+ m^2 \right)\W(x_1,x_2)=\left(\square_{x_2}
 + m^2 \right)\W(x_1,x_2)
= 0\, . \label{kkgg}
\end{equation}
Because of  Eq. (\ref{poi1}), $\W(x_1, x_2)$ is also required to be a solution of the functional equation
\begin{equation}\label{maincondition}
\W(x_1, x_2)- \W(x_2,x_1) = C(x_1,x_2)
\end{equation}
in the sense of distributions. 

Starting from $\W(x_1, x_2)$, the Hilbert space of the theory ${\cal H}$ can be constructed
by using standard techniques \cite{pct}.
The first, step consists in giving a norm to the one-particle state $\Psi_f$  corresponding to a given test function $f\in \T(\M)$; the norm is computed by using the  two-point function:
\begin{equation}\label{scp}
||\Psi_f||^2= \int_{\M\times \M}  \W(x_1, x_2)f^*(x_1)f(x_2) \,    \sqrt{-g(x_1)}dx_1\, \sqrt{-g(x_2)}dx_2. 
\end{equation}
 The squared norm (\ref{scp}) is positive (as it should) if $\W(x_1,x_2)$  satisfies the {\em positive-definiteness condition} which is nothing but the nonnegativity ot the rhs of Eq. (\ref{scp}). We assume that it does.

The norm (\ref{scp}) actually  comes from  a pre-Hilbert scalar product
whose interpretation is that of providing the quantum transition amplitudes between two one-particle states: 
\begin{equation}\label{scp2}
\langle\Psi_f,\Psi_g\rangle = \int_{M_d}\W(x_1,x_2)  f^*(x_1) g(x_2)\,  \sqrt{-g(x_1)}dx_1, \sqrt{-g(x_2)}dx_2. 
\end{equation}
The one-particle Hilbert space  ${\cal H}^{(1)}$ is obtained by quotienting the  subspace  of zero norm states and by taking the  Hilbert completion. 
The full Hilbert space is the symmetric Fock space
$${\cal H} = F_s({\cal H}^{(1)})={\cal H}_0 \oplus [ \oplus_n Sym({\cal H}_1)^{\otimes n}]$$
(with $Sym$
denoting the symmetrization operation and  ${\cal H}_0 =\{\lambda 1,
\lambda \in {{\Bbb C}}\}$ ).
In the final step one introduces the field operator $\widehat\phi(f)$ decomposed into its 
``creation'' and ``annihilation'' parts; 
\begin{equation}
\widehat\phi(f)=\widehat\phi^{+}(f)+\widehat\phi^{-}(f); \label{azzone}
\end{equation}
the latter are 
defined by their action on the dense subspace ${\cal H} ^{(0)}$ of vectors having finitely many non-vanishing components: ${\Psi} = (\Psi_0, \Psi_1,\ldots
\Psi_k,\ldots,0, 0,0,\ldots)$:
\begin{align}
&\left({\widehat{\phi}}^{-}(f){\Psi}\right)_{n}=
{\sqrt{n+1}}\int {\W}(x,x') f(x)
{\Psi}_{n+1}(x',x_1,\ldots,x_n)
 \sqrt{-g(x)}dx \sqrt{-g(x')}dx',
\\
&\left(\widehat\phi^{+}(f){\Psi}\right)_n=\frac{1}{\sqrt{n}}
\sum_{j=1}^{n}f(x_j){\Psi}_{n-1}(x_1,\ldots, \slashed{x}_{j},\ldots,x_n).
\label{azzo}
\end{align}
Eq.  (\ref{maincondition}) shows that these formulae do imply the commutation relations (\ref{poi1}) and that
\begin{equation}
\W(x,x') = \langle \Psi_0, \widehat \phi(x) \widehat \phi(x') \Psi_0 \rangle 
\end{equation}
where
\begin{equation}{\Psi_0} = (1, 0,0,\ldots,)
\end{equation}
is the cyclic reference state of the representation.

In the end, either in flat or  curved spacetime, quantizing a free field theory
amounts to specify its  two-point function which carries all the information about the Hilbert space and the field operators. 
Furthermore, the knowledge
of the two-point function and the commutator allow to determine the  Green's functions modulo the necessary renormalizations;  the two-point function  therefore encodes
not only the dynamics of the free field but also 
the possibility of studying interactions  perturbatively. 

But, how do we specify a criterion to choose among the infinitely many existing possibilities? Here comes the spectral condition.

\section{Prelude: the Spectral condition in Minkowski space} \label{mink}
This section contains material that may be found in (good) textbooks. The reason to recall it here is to better appreciate and understand the role of the spectral condition and plane waves in the de Sitter and anti de Sitter contexts.

 On page 97 of the classic book by R. Streater and A,  S. Wightman, the following basic assumption about a relativistic quantum field theory is declared:
\begin{quote}
{\em Axiom 0. Assumptions of Relativistic Quantum Theory.}

The states of the theory are described by unit rays in a separable Hilbert space ${\cal H}$. The relativistic transformation law of the states is given by a continuous unitary representation of the inhomogeneous Lorentz group  $\{a,A\}\rightarrow U(a,A)$. Since $U(a,1)$ is unitary, it can be written as $U(a,1)= \exp(i a_\mu P^\mu)$ where $P^\mu$ is an unbounded operator interpreted as the energy momentum operator of the theory. The eigenvalues of $P^\mu$ is lie in or on the forward cone ({\em spectral condition}). There is an invariant state $\Psi_0$, \ \  $U(a,1)\Psi_0 = \Psi_0$  unique up to a constant phase factor ({\em uniqueness of the vacuum}).
\end{quote} 
Stated more succintly:
\begin{quote}
The joint spectrum of the infinitesimal generators of $U(a,1)$ 
lies in the closed forward cone $\overline V_+$.
\end{quote}
This is the spectral condition of standard (zero temperature) QFT.  It is its most important and characteristic feature. All the other axioms are of a kinematical character\footnote{Apart from the nonlinear (and hard to verify) positivity condition of the correlation functions necessary to reconstruct a Hilbert space.}.

Here we consider a general $d$-dimensional Minkowski spacetime $M_{d}$ with metric 
\begin{equation}
\eta_{\mu\nu}= diag(+,-,\dots,-)
\end{equation}
 and one scalar field. The open future cone  of the origin (also called the forward cone)  is the set
\begin{eqnarray}
V_+ &=& \{x\in M_{d}\ :\ x\cdot x >0,\ \ \ x^0> 0\}.
\end{eqnarray}
Given the $n$-point vacuum expectation values of the field (in short: the $n$-point functions): 
\begin{equation} 
\W_n(x_1,\ldots x_n) = \langle \Psi_0,\widehat \phi(x_1)\ldots\widehat\phi(x_n)\Psi_0\rangle,
\end{equation}
the spectral condition is immediately translated into a property of the support of their Fourier transforms $\widetilde W_n(p_1,\ldots,p_n) $: the distribution 
\begin{equation} 
 \widetilde \W_n (p_1,\ldots,p_n) = \int e^{i p_1\cdot x_1+\ldots +i p_n\cdot x_n }\W_n(x_1,\ldots ,x_n) dx_1\ldots dx_n
 \end{equation} 
vanishes unless all momenta are in the energy-momentum spectrum of the states: 
\begin{equation} 
p_1\in \overline V_+, \ \ \  p_1+p_2 \in \overline V_+,  \ldots \  \  \ p_1+p_2+\ldots +p_n \in \overline V_+ .
 \end{equation} 
By Fourier-Laplace transform,  support properties in one space are give rise to analyticity properties in the dual space \cite{pct}. 
A fundamental theorem of this category shows  that the $n$-point distributions are boundary values of  $n$-point functions holomorphic in tubular in domains of the complex  Minkowski spacetime:
\begin{theorem}[A. S. Wightman]
the distribution $\W_n(x_1,\ldots x_n)$  is the boundary value of a function $W_n(z_1,\ldots z_n)$ holomorphic in the  tube 
\begin{equation} {T}_n= \{(z_1,\ldots z_n) \ : \ \ \Im (z_{j+1}-z_j ) \in \overline V_+\}.
\end{equation} 
\end{theorem}
Wightman's reconstruction theorem \cite{pct} finally  states the {\em equivalence of the analyticity of the $n$-point function  in the tubes $T_n$ and the spectral condition}:  starting from a set of Wightman  functions having such analyticity properties, it is possible to reconstruct the Hilbert space of the theory, the representation of the inhomogeneous Lorentz group, the infinitesimal generators of the translation group and verify that their joint spectrum is contained in the closed forward cone. 

The above analyticity properties and the spectral condition have therefore one and the  very same precise {\em physical meaning}: they assert  that the states  of the theory have positive energy  in every Lorentz frame. 

\vskip 10 pt 

Focusing now on  two point functions, the spectral condition is equivalent to the following simpler  property:
\begin{Corollary}[Normal analyticity property]
$\W(x_1,x_2)$  is the boundary value of a function $W(z_1, z_2)$ holomorphic in the tube $ T_{12}= {T}_-\times {T}_+$: 
\begin{equation}
\W(x_1,x_2) = \langle \Psi_0, \widehat\phi(x_2)\widehat  \phi(x_1)\Psi_0 \rangle =\underset{\underset{  T_+ \ni \, z_2 \to x_2 }{\scriptscriptstyle  T_- \ni \, z_1\to x_1 }}{ b.v.} {{ W}}(z_1,z_2)
\end{equation} 
where 
\begin{equation}
{T}_\pm= \{(z = x+i y : \ \ \ \pm y \in \overline V_+\}
\end{equation}
are the past and future tubes.
\end{Corollary}
The tubes $T_\pm$ are invariant under the action of the real inhomogeneous Lorentz group. Acting with the complex group one discovers that every Lorentz invariant  two-point function satisfying the spectral condition enjoys a much larger analyticity domain:

\begin{theorem} [Maximal analyticity property]
\begin{enumerate}\item The two-point function $W(z_1,z_2)$  depends only on the Lorentz-invariant  variable $\lambda= (z_1-z_2)^2$. 

\item $W(z_1,z_2)$ can be continued to a function $ {\frak W}(z_1,z_2)$ analytic in the cut-domain
\begin{equation}
\Delta_0 =  \{ (z_1,z_2 ); \ \ (z_1-z_2)^2\not = \rho ,\ \   \rho \geq0\}
\end{equation}
which contains all pairs of complex events with the exception of all pairs of real events which are  causally connected (the causal cut).
 
\item $ {\frak W}(z_1,z_2)$ is invariant in $\Delta_0 $ under the action of the complex inhomogeneous Lorentz group. 

\item The permuted two-point function
is the boundary value of  ${\frak W}(z_1,z_2)$
from the opposite tube $
{ T}_{21} = T_+\times T_-$:
\begin{equation}
\W(x_2,x_1) = \langle \Psi_0, \widehat\phi(x_2)\widehat  \phi(x_1)\Psi_0 \rangle =\underset{\underset{  T_-\ni \, z_2 \to x_2 }{\scriptscriptstyle  T_+ \ni \, z_1\to x_1 }}{ b.v.} {\frak{ W}}(z_1,z_2).
\end{equation} 
\item  The cut-domain $\Delta_0$  contains all  pairs of non-coinciding Euclidean points 
\begin{equation}
\dot{\cal E}= \{z_1, z_2 \in \Delta, \ \ \Re z_1^0=\Re z_2^0 = 0, \ \Im z_1^i =\Im z_2^i =  0 , \ i=1,\ldots,d-1 ,\ \  z_1\not=z_2 \}.
\end{equation}
 The Schwinger function $S$ (also called  the Euclidean propagator) is the restriction of ${\frak W}(z_1,z_2)$ to the non-coincident Euclidean points
 $\dot{\cal E}$.  $S$ is  analytic in  $\dot{\cal E}$ and can be extended as a distribution to the whole Euclidean space ${\cal E}$ including the coinciding points.
\end{enumerate}
\label{maxan}
\end{theorem}

\subsection{Klein-Gordon fields}
Let us see now how the spectral condition works in practice for Klein-Gordon fields. The first thing is to identify 
a suitable basis of solutions of the Klein-Gordon operator. 
In flat space the exponential plane waves are almost always the convenient choice, since they are also characters of the translation group:
\begin{equation}
\psi^{(\pm)}_{\vec p} (x) = \frac 1{2\sqrt {(2\pi)^{d-1} \omega}} \exp(\pm i p x), \ \ \ p^0=\omega=\sqrt{|\vec{p}|^2+m^2}.
\end{equation}
The above plane waves extend to functions that are holomorphic in the whole complex Minkowski spacetime $M_d^{(c)}$. 
The important point to be noticed is the following:
\begin{Remark} \label{remark1}
Positive frequency waves $ \psi^{(-)}_{\vec p} (z)$ are exponentially decreasing in the past tube $T_-$;  negative frequency waves $ \psi^{(-)}_{\vec p} (z)$  are exponentially decreasing in the future tube  $T_+$.
\end{Remark}

Let us now examine the two-point function.  By translation invariance it may depend only on the difference variable  $\xi= x_1-x_2$.
Taking the Fourier transform of the Klein-Gordon equation w.r.t $\xi$ gives
\begin{equation}
(p^2-m^2)\tilde \W_m(p)=0.
\end{equation} 
The most general Lorentz invariant distributional solution has two disconnected components:
\begin{equation}
\tilde \W_m(p)= a\theta(p^0) \delta(p^2-m^2) +  b\theta(-p^0) \delta(p^2-m^2) 
\end{equation}
and the spectral condition imposes  $b=0$.
By Fourier anti-transforming  we get :
\begin{eqnarray}\label{tp}
\W(x_1,x_2) =  \frac{1}{2(2\pi)^{d-1}}
\int 
\frac{e^{-i\omega(x_1^0-x_2^0) + i\vec{p}\cdot(\vec{x_1}-\vec{x_2})}}{\sqrt{|\vec p|^2 + m^2}}\,d\vec p = 
\int  \psi^{(-)}_{\vec p} (x_1)  \psi^{(-)}_{\vec p} (x_2)d\vec p . 
\end{eqnarray}
Remark \ref{remark1} invite us to move the first point into the past tube  $T_-$ and the second point into forward tube $T_+$; 
this move greatly improves the convergence of the integral: the function 
\begin{eqnarray}
W_m(z_1,z_2) =  
\int  \psi^{(-)}_{\vec p} (z_1)  \psi^{(+)}_{\vec p} (z_2)d\vec p ,  \ \ \ z_1\in T_, \ \ (z_1,z_2) \in  T_+ \label{naflat}
\end{eqnarray}
is now an analytic function of $(z_1,z_2) \in T_-\times T_+$. The two-point distribution $ \W(x_1,x_2)$ is recovered by taking the boundary value. 
The normalisation ensures that the CCR's hold with the correct
coefficient. 

\vskip 10 pt
Let us discuss the  elementary massless case in more detail: 
\begin{eqnarray}\label{tp1}
 W((t-is,0\ldots,0),0)& =&\int
 \frac{e^{-i\omega(t-is)}{k}^{d-3}}{2(2\pi)^{d-1}} e^{-i\omega(t-is)}{k}^{d-3}\,dk \
d\Omega_{d-2} \ 
\\ &=&  \frac{1}{(4\pi)^\frac{d-1}{2}}
\frac{\Gamma(d-2)}{\Gamma\left(\frac{d-1}{2}\right)}\frac{1}{(it+s)^{d-2}}.
\end{eqnarray}
By restoring in this expression the Lorentz-invariant variable
$(z_1-z_2)^2$ we get right away  the maximally analytic two-point function:
\begin{equation}\label{tpx}
{\frak W}(z_1,z_2) =
\frac{\Gamma\left(\frac{d-2}{2}\right)}{4\pi^\frac{d}{2}}[-(z_1-z_2)^2]^{-\frac{d-2}{2}}.
\end{equation}
Its boundary values from the relevant tubes give the two-point function $\W(x_1,x_2)$ and the permuted two-point function $\W(x_2,x_1)$.
The covariant commutator  is their difference (\ref{maincondition}):
\begin{equation}\label{CCR}
C(x,y) = \frac{\Gamma\left(\frac{d-2}{2}\right)}{4\pi^\frac{d}{2}}
\left([-(x-y)^2 + i \varepsilon (x^0-y^0)]^{-\frac{d-2}{2}} -
[-(x-y)^2 - i \varepsilon (x^0-y^0)]^{-\frac{d-2}{2}}\right).
\end{equation}
Using the notations of \cite{gelfand} we obtain that
\begin{equation}\label{CCR2}
C(x,y) = \frac{1}{2\pi
i}\frac{1}{\Gamma\left(2-\frac{d}{2}\right)}\varepsilon(x^0-y^0)
[-(x-y)^2]^{-\frac{d-2}2}_-.
\end{equation}
where $\varepsilon(x) = \theta(x)-\theta(-x)$. 
When the spacetime dimension 
$d$ is even the distribution $[-(x-y)^2]^{\lambda}_-$ has a simple
pole at $\lambda = -\frac {d-2}2 $ with residue
\begin{equation}
{\mathrm {Res}}_{\lambda = -\frac{d-2}{2}} [-(x-y)^2]^{\lambda}_-
=\frac{(-1)^{\frac d2 -2}}{\Gamma\left(\frac d2 -
2\right)}\delta^{(\frac d2 -2)}[(x-y)^2]
\end{equation}
while $1/{\Gamma\left(\frac d2 - 2\right)}$ has a zero and we obtain that the support of the commutator is the light-cone (Huygens principle):
\begin{equation}\label{CCR3}
C(x,y) = \frac{1}{2\pi i}\varepsilon(x^0-y^0)\delta^{(\frac d2
-2)}[(x-y)^2].
\end{equation}
In particular for $d=4$ we get the well-known dominant term of the
Pauli-Jordan function.

\newpage
\section{de Sitter}
Let us consider now  the $d$-dimensional de Sitter universe (see Eq. (\ref{dsman})):
\beq
dS_d = \{x \in M_{d+1}\ :\ x\cdot x = -R^2=-1\}\ .
\label{dsdd}\endq
The future cone of the origin of the ambient spacetime in one dimension more 
\beq
V^+= \{x \in M_{d+1}\ :\ x ^2>0 ,  \ x^0>0\}
\endq
provides the causal ordering of  the de Sitter manifold. An event $x_2$ is in the future of another event $x_1$ if the vector $(x_2-x_1)$ belongs to the closed future cone $\overline V^+$ of the ambient spacetime. 
Two events $x_1, x_2 \in dS_d$ are spacelike separated if and only if 
\begin{equation}
(x_1-x_2)^2= -2 -2x_1\cdot x_2 <0. \label{causal}
\end{equation}

A straightforward adaptation of the spectral condition of Wightman QFT is just not possible because there exists no global energy operator available in the de Sitter case.  This is a consequence of the absence of a global timelike Killing vector field on the de Sitter manifold. Timelike Killing vector fields exist only on wedge-like submanifolds bordered by bifurcate Killing horizons,  but there is no Killing vector field that remains timelike on the whole manifold.

Still, since the complexification of Minkowski space plays such a crucial role in Minkowski QFT, we may go on and consider the complex de Sitter space-time,  here visualised as a submanifold of the complex $(d+1)$-dimensional Minkowski space:
\beq
dS_d^{(c)} = \{z \in M_{d+1}^{(c)}\ :\ z\cdot z = -R^2=-1\}\ .
\label{s.2}\endq
Note that $z=x+iy \in dS_d^{(c)}$ if and only if $x^2-y^2=-R^2$ and $x\cdot y = 0$. 
On the complex manifold there acts the complex de Sitter group $G^{(c)}$, which is the complexification of the restricted Lorentz group of the ambient space $G= SO_0(1,d)$. 

$dS_d^{(c)}$  contains tuboids $\TT_\pm$ which are closely similar to the past and future tubes of Minkowski space: actually, they are described in the simplest way precisely as  the intersections of the ambient tubes $T_\pm$ with the complex de Sitter manifold:
\beq
\TT_\pm =   dS_d^{(c)} \cap T_\pm= \{x+iy \in dS_d^{(c)}\ :\ y\in \pm V_+\}\ .
\label{s.2.1}\endq
The set of points with purely imaginary zero component $z^0= i y^0$ and purely real spatial components  $z^i= x^i, \ i = 1,\ldots,d$, is the Euclidean sphere of the complex de Sitter manifold:
\begin{equation}
S_d= \{z = (iy^0, x^1,\ldots,x^d)
\in {\bf C}^{1+d} : {y^0}^2 +{x^1}^2+\ldots +{x^d}^2=R^2=1 \}. \label{cds}
\end{equation}

Let us come to  de Sitter QFT.   
While it is impossible to formulate a true spectral condition, we may retain its most characteristic consequence: in the case of two-point functions we may  assume   \cite{bm}  that 
there holds the following \vskip 10 pt
\noindent {\bf Assumption 1:  Normal analyticity property}. {\em  
$\W(x_1,x_2)$   is the boundary value of a function $W(z_1, z_2)$ holomorphic in the tube $ {\cal T}_{12}= {\TT}_-\times {\TT}_+$, 
\begin{equation}
\W(x_1,x_2) = \langle \Psi_0, \widehat\phi(x_2)\widehat  \phi(x_1)\Psi_0 \rangle =\underset{\underset{  \TT_+ \ni \, z_2 \to x_2 }{\scriptscriptstyle  \TT_- \ni \, z_1\to x_1 }}{ b.v.} {{ W}}(z_1,z_2)
\end{equation} 
where 
$
{\TT}_- $ and $
{\TT}_+$
are the de Sitter past and future tubes.}
\vskip 10 pt
Of course the physical interpretation  of this property cannot be the positivity of the energy spectrum of the states. 
It turns out that the correct physical interpretation is thermodynamical \cite{bm,hawking,bemds}.
\begin{figure}[h]
\includegraphics[width=14cm]{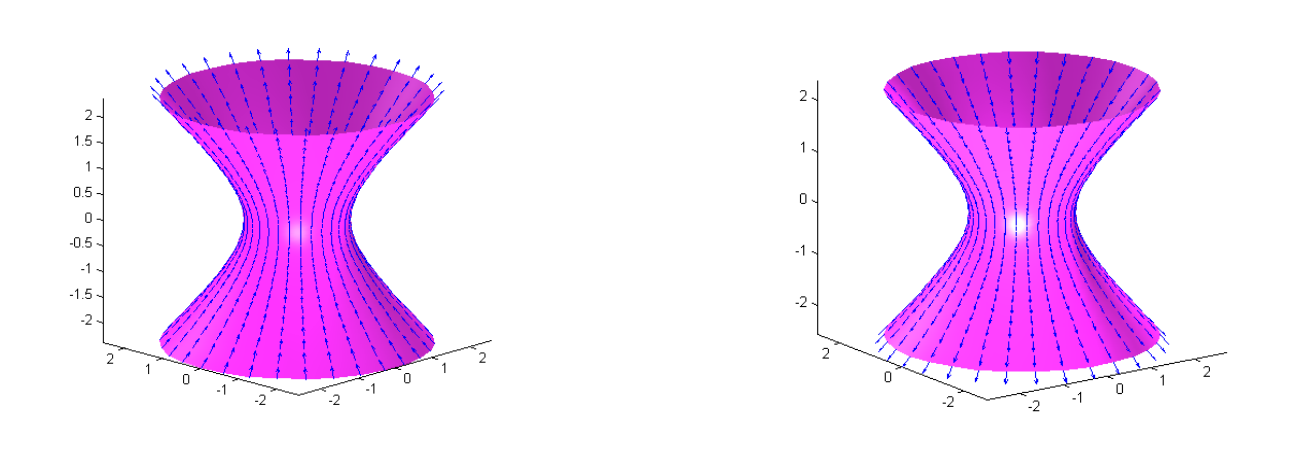} 
\caption{Sections of the forward and the backward  tubes in the complex dS manifold. 
The arrows are the imaginary parts $y$ of the vectors $z=x+i y$, represented as attached at the endpoint of their real parts $x$ which belong to  the hyperboloid whose radius is $-R^2+y^2$. Here is represented the section at a given fixed positive value of $y^2$. Recall that $x\cdot y =0$. \label{fig3b}}
\end{figure}   

The tubes $\TT_\pm$ are invariant under the action of the real de Sitter group. By acting with the complex group, a much larger analyticity domain appears as before. 
The following theorem  \cite{bm} is {\em mutatis mutandis} identical to theorem \ref{maxan} \cite{bm}: 
 \begin{theorem} [Maximal analyticity property]
\begin{enumerate}\item The two-point function $W(z_1,z_2)$  depends only on the Lorentz-invariant  variable $\zeta = z_1\cdot z_2$.
\item $W(z_1,z_2)$ can be continued to a function $ {\frak W}(z_1,z_2)$ analytic in the cut-domain
\begin{equation}
\Delta =  \{ (z_1,z_2 ); \ \ \zeta \not= \rho ,\ \   \rho \leq -1\}
\end{equation}
which contains all pair of complex events minus the causal cut (\ref{causal}).
 
\item $ {\frak W}(z_1,z_2)$ is invariant in under the action of the complex de Sitter group. 

\item The permuted two-point function
is the boundary value of  ${\frak W}(z_1,z_2)$
from the opposite tube $
{ T}_{21} = T_+\times T_-$:
\begin{equation}
W(x_2,x_1) = \langle \Psi_0, \widehat\phi(x_2)\widehat  \phi(x_1)\Psi_0 \rangle =\underset{\underset{  T_-\ni \, z_2 \to x_2 }{\scriptscriptstyle  T_+ \ni \, z_1\to x_1 }}{ b.v.} {\frak{ W}}(z_1,z_2).
\end{equation} 
\item  The cut-domain $\Delta$  contains the all the non-coinciding Euclidean points 
\begin{equation}
\dot{\cal E}= \{z_1, z_2 \in \Delta, \ \ z_1 \in S_d , \  z_2 \in S_d , \ \ z_1\not=z_2 \}.
\end{equation}
The Schwinger function $S$  is the restriction of ${\frak W}(z_1,z_2)$ to the non-coincident Euclidean points
 $\dot{\cal E}$.  $S$ is  analytic in  $\dot{\cal E}$ and can be extended as a distribution to the whole Euclidean space ${\cal E}$ including the coinciding points.
\end{enumerate} 
\end{theorem}

\subsection{Klein-Gordon fields and plane waves}
Now we want to construct  dS Klein-Gordon quantum fields  starting from two-point functions (as in Eq. (\ref{azzone})) that are {\em normal analytic} in the sense of Assumption 1. Following the paradigm of flat space,   
we should look for wave solutions of the Klein-Gordon equation analytic in the past and, respectively,  the future tube 
and write a two-point function similar to Eq. (\ref{naflat}).
When solving the Klein-Gordon equation the normal strategy is separating the variables; however this  would not be a good idea if the normal analyticity property has to appear manifestly as in Eq. (\ref{naflat}). 

\begin{figure}[h]
\includegraphics[width=10.5 cm]{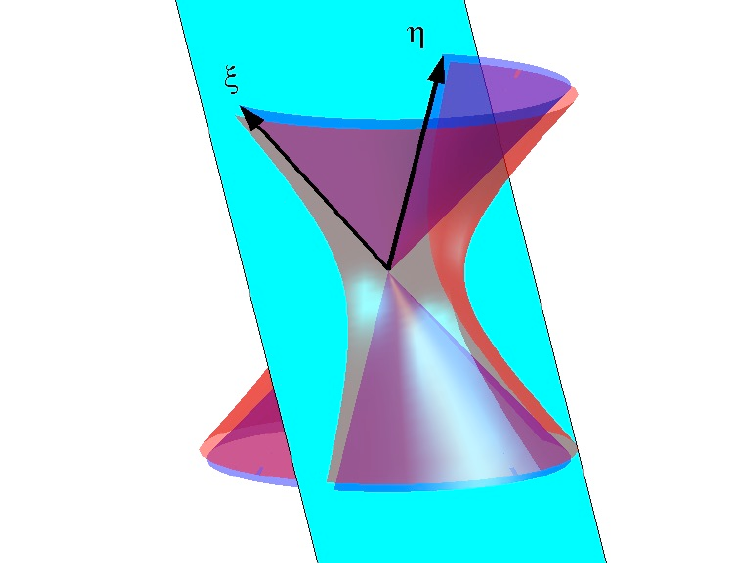}
\caption{Timelike geodesics can be parametrized by the choice of two null vectors in the ambient space; they have the physical interpretation of asymptotic momentum directions.\label{fig1}}
\end{figure}

One possibility  comes from the study of geodesics \cite{geo}:  a de Sitter  timelike geodesics may be parametrized by the choice of two lightlike vectors belonging to the future lightcone $C_+ $ of the ambient Minkowski spacetime (see Fig. \ref{fig1}) as follows:
\begin{eqnarray}
x^\mu(\tau) = \frac R {\sqrt{2 \xi\cdot \eta}} ( \xi^\mu e^{\frac\tau R} - \eta^\mu e^{-\frac \tau R}). \label{geod}
\end{eqnarray}
The two null vectors parametrizing the geodesics point towards its asymptotic directions.  In fact, the  conformal compactification of the Sitter manifold has a boundary at timelike infinity and the lightcone of the ambient spacetime is in a precise sense equivalent to it \cite{Bros:2001yw}.
A natural  basis of 
 solutions of the de Sitter Klein-Gordon equation
\begin{equation} \Box_{dS} \psi (z)   +m^2 \psi (z) =0 \label{KGnu}\end{equation}
may be thus 
parametrized by the choice of a lightlike vector $\xi\in C^+$  and a complex number $\lambda$  as follows:
\begin{equation}
\psi_\lambda(z,\xi) = \left({ z\cdot \xi}\right)^{\lambda}  = e^{\lambda \log \left({ z\cdot \xi}\right) }   \label{waves}.
\end{equation}
In this definition the scalar product is in the sense of the ambient spacetime. 
The  functions  (\ref{waves}) are plane waves as their phase is constant on the  planes $z\cdot \xi = const$.  
As required, they are well-defined and analytic in each of the tubes
$\cal T^+$ and $\cal T^-$ \cite{bm}. 

It useful to introduce a new complex parameter $\nu$ by the following definition:
\begin{equation}
\lambda= 
-\frac{d-1}2+i \nu.
\end{equation}
The parameters $\lambda$ and $\nu$ are related to the complex mass squared and the complex dimension as follows:
\begin{equation}
m^2 = -\lambda(\lambda+d-1) = \frac {(d-1)^2}{4} +\nu^2. \label{cmass}
\end{equation}
Of course $m^2$ is real and positive only when: 
\begin{enumerate}
\item $\nu$  is real; this correspond in a group-theoretical language to the principal series of unitary representations of the Lorentz group; 

\item $\nu$ is purely imaginary and  $|\nu|<\frac{d-1}{2} $; this correspond to the complementary series of unitary representations of the Lorentz group.
\end{enumerate}
But in the de Sitter universe there is room also for negative mass squared at certain discrete values \cite{tak1,tak2}.
\subsection{Construction of the two-point function}
We may now mimick Eq. (\ref{naflat})  and   consider the two-point function:
\begin{eqnarray}
 \int_{\gamma}(\xi\cdot z_1)^{-\frac{d-1}2 - i \nu}(\xi\cdot z_2)^{-\frac{d-1}2 + i \nu}\, 
d\mu_\gamma, \ \ \  z_1 \in \TT_-, \   \ \ z_2 \in \TT_+
\label{r.2} \end{eqnarray}
where 
\begin{equation} d\mu_\gamma(\xi)= \alpha(\xi)|_\gamma = (\xi^0)^{-1} \sum_{j=1}^d (-1)^{j+1}
\xi^j\,d\xi^1\ldots\ \wh{d\xi^j}\ldots\ d\xi^d|_\gamma\ .
\label{f.16}\end{equation}
and $\gamma$
denotes any
($d-1$)-dimensional integration cycle in  $C^+$. 
To fix the ideas we may integrate over the spherical basis  $\sd$ of the cone $C^+$ equipped with its canonical orientation:
\begin{equation}
 \gamma_0=\sd= C^+ \cap \{\xi\ :\ \xi^0 =1\}
=  \{ \xi \in C^+ : {\xi^1}^2 + \ldots + {\xi^d}^2 =
1\}.
\end{equation}
In this case $\alpha(\xi)|_\gamma$  coincides
with the rotation invariant measure $d\mu_{\gamma_0}$ on  $\sd$ normalized as
follows:
\begin{equation}
\omega_{d}=\int_{\gamma_0}d\mu_{\gamma_0}  = \frac{2\pi^\frac d2}{\Gamma\left(\frac d2\right)}.
\label{norms}
\end{equation}
It is self-evident that 
\begin{property} 
The two-point function (\ref{r.2})  solves by construction the Klein-Gordon equation in each variable and is manifestly holomorphic in $\TT_-\times\TT_+$.
\end{property} 
Since the integrand is a homogeneous function of $\xi$ of degree $(1-d)$, the integral (\ref{r.2}) is actually  the integral of a closed differential form and, as such, does not depend on the integration cycle. This immediately implies that 
\begin{property} 
The two-point function (\ref{r.2}) is de Sitter invariant and depends only on the invariant $\zeta = z_1\cdot z_2$.
\end{property} 
To compute it explicitly   we may therefore choose  the two arbitrary points  $z_1$  in $ \TT^-$ and $z_2$  in $ \TT^+$ in the way  the most pleases us. 
Interestingly, different choices produce different integral representations of the same function. 
A useful choice is
$$z_1=(
-i ,
0 ,
\ldots,
0,
0 ),
 \  \ \ z_2(s)= (
\sinh(i s) ,
0 ,
\ldots, 0,
\cosh( is)),
\ \  \
\xi = \left(
 1 ,
\vec{n}   \sin \theta,
 \cos \theta
\right),
$$
so that
\begin{equation}
\zeta = z_1\cdot z_2(s)=  \sin (s ), \ \ (\zeta^2-1)^\frac 12  = i \cos s.
\end{equation}
The condition  $z_2$  in $ \TT^+$ means  $0<s<\pi$. We get \cite[Eq. 7, p 156]{bateman}
\begin{eqnarray} 
&& \int_{\sd}(\xi\cdot z_1)^{-\frac{d-1}2 - i \nu}(\xi\cdot z_2)^{-\frac{d-1}2 + i \nu}\,
d\xi = \cr &=& \omega_{d-1} \int_{0}^{\pi}e^{-\frac{i\pi}{2}(-\frac{d-1}2 - i \nu)}(i \sin s - \cos s \cos\theta )^{-\frac{d-1}2 + i \nu}\, \sin^{d-2}\theta 
d\theta
\cr
\cr &=& (2\pi)^{{d \over 2}}  e^{-\pi \nu}   (\zeta^2-1)^{-{d-2 \over 4}}\,P_{-\frac 12 + i \nu}^{-{d-2 \over 2}}(\zeta).
\label{rppp} \end{eqnarray}
\vskip 10 pt

\noindent Imposing the normalization of the CCR's gives the plane-wave expansion of the two-point function, valid for any complex value of $\nu$ that is not a pole of $\Gamma\left(\frac{d-1}2 + i \nu\right)\Gamma\left(\frac{d-1}2 - i \nu\right)$:\\

\noindent {\bf Main formula}: 
The canonically normalized (so-called Bunch-Davis) Wightman function of a Klein-Gordon de Sitter scalar field has the following expressions:
\begin{eqnarray}
&& W_\nu (z_1,z_2)   = w_\nu(z_1\cdot z_2) = 
\cr&&
\cr && =  {\Gamma\left(\frac{d-1}2 + i \nu\right)\Gamma\left(\frac{d-1}2 - i \nu\right)
e^{\pi\nu}\over 2^{d+1}\pi^d} 
\int_{\gamma}(\xi\cdot z_1)^{-\frac{d-1}2 - i \nu}(\xi\cdot z_2)^{-\frac{d-1}2 + i \nu}\,\formdif(\xi)    \label{uup} \\ && =  {\Gamma\left(\frac{d-1}2 + i \nu\right)\Gamma\left(\frac{d-1}2 - i \nu\right)   \over 2 (2\pi)^{d/2}}
(\zeta^2-1)^{-{d-2 \over 4}}\,P_{ -{1\over 2}+ i \nu }^{-{d-2 \over 2}}(\zeta).
\label{wig}
\end{eqnarray}
 Eq. (\ref{uup}) is  only valid in the normal domain of analyticity $z_1$  in $ \TT^-$ and $z_2$  in $ \TT^+$; on the other hand  the rhs of  equation (\ref{wig}) is maximal analytic that is entire in the cut-plane $\Delta $.

\vskip 10pt

The discontinuity of the Wightman function on the cut provides the commutator; the retarded propagator  function is obtained by (carefully) multiplying the commutator with the relevant step function:
\begin{eqnarray}
&&C_\nu(x_1,x_2)  = \W_\nu(x_1,x_2)- \W_\nu(x_2,x_1), \\
&&R_\nu(x_1,x_2)  = i  \theta(x_2,x_1) C_\nu(x_1,x_2).  \label{appret}
\end{eqnarray}
To compute the retarded propagator let us choose $x_2$  in the future cone of the origin: 
$$x_0=(0,0, \ldots,0,1)
 \ \ \  x_2(t)= (
\sinh t ,
0 ,\ldots , 0
\cosh t ),
\ \ \ 
t>0, \ \ \ \zeta =-\cosh t .
$$
%
The retarded discontinuity $(x_2>x_1)$ is therefore
\begin{eqnarray}
 r_\nu(u) =   {i \, \Gamma\left(\frac{d-1}2 + i \nu\right)\Gamma\left(\frac{d-1}2 - i \nu\right)   \over 2 (2\pi)^{d/2}}
(u^2-1)^{-{d-2 \over 4}}\,\left(P_{ -{1\over 2}+ i \nu }^{-{d-2 \over 2}}(\zeta-i\epsilon)-P_{ -{1\over 2}+ i \nu }^{-{d-2 \over 2}}(\zeta+i\epsilon) \right) \cr 
=  \cosh\left(\pi \nu\right)
{\Gamma\left(\frac{d-1}2 + i \nu\right)\Gamma\left(\frac{d-1}2 - i \nu\right)   \over  (2\pi)^{d/2}}
(\zeta^2-1)^{-{d-2 \over 4}}\
P_{-{1\over 2}+i\nu}^{-{d-2 \over 2}}(-\zeta) .\cr
\label{retprop}
\end{eqnarray}
The Schwinger function is the restriction of the maximally analytic two-point function to the Euclidean sphere. Given any two points of the Euclidean sphere their invariant product may be parametrized as follows
$ z_1\cdot 
z_2 = -\cos(s)$; the choice of sign is because at coincident points $z^2=-1$. Thus
\begin{eqnarray}
 G_{\nu}(- \cos s ) = {\Gamma({{d-1\over 2} + i\nu} )\Gamma({{d-1\over 2} - i\nu} )\over 2 (2\pi)^{d/2}} (\sin s)^{-{d-2 \over 2}}\,e^{\frac{i\pi}2(d-2)}\P_{-{1\over 2} +i \nu }^{-{d-2 \over 2}}(-\cos s).  \label{leg6}
\end{eqnarray}
At this point we are fully equipped to begin studying perturbative quantum field theory on the the de Sitter universe. 
Of course we do not do it here but we want to discuss one remarkable success of the above formalism.

\subsection{Linearization and the K\"all\'en-Lehmann representation}
In Minkowski space, any scalar  two-point function $W(z_1,z_2)$  satisfying the properties described in Sect. \ref{mink}  admits a K\"all\'en-Lehmann representation of the form 
\beq 
W(z_1, z_2) = \int_0^\infty \rho (m^2)  \, W_m(z_1, z_2)\,  dm^2 
\endq
 where $W_m(z_1, z_2)$ is given in Eq. (\ref{naflat}) 
 and the  weight $\rho (m^2)$ is a positive measure of tempered growth. In particular,  given
two masses $m_1$ and $m_2$, computing the weight for the bubble 
\beq 
W_{m_1}(z_1, z_2) W_{m_2}(z_1, z_2) = \int_{(m_1+m_2)^2}^\infty \rho (m^2:m_1,m_2)  W_m(z_1, z_2) dm^2
\endq 
is an easy exercise of Fourier transformation. 

The corresponding de Sitter case is much more difficult; to obtain the K\"all\'en-Lehmann  weight of the corresponding integral
 \begin{eqnarray}
W_\lambda(z_1,z_2)W_\nu(z_1,z_2) =\int_{-\infty}^{\infty}  \rho({\lambda},\nu,\kappa) W_\kappa (z_1,z_2) \kappa d\kappa,
\label{KL}
\end{eqnarray}
one should compute the Mehler-Fock transform of $W_\lambda(\zeta)W_\nu(\zeta) $; this amounts to  the following integral:
\begin{eqnarray}
h_d(\lambda,\nu, \n) = \int_1^\infty
{P^{-\frac{d-2}{2}}_{-\frac{1}{2} +
i\lambda}(u)P^{-\frac{d-2}{2}}_{-\frac{1}{2} +
i\nu}(u)} P^{-\frac{d-2}{2}}_{-\frac{1}{2} + i \n} (u)\,
{(u^2-1)^{-\frac{d-2}4}} \ du \label{hdn}
\end{eqnarray}
and the K\"all\'en-Lehmann  weight is 
\begin{eqnarray}
 \rho({\lambda},\nu,\kappa) = \frac{\Gamma\left(\frac{d-1}{2}+i \nu\right)\Gamma\left(\frac{d-1}{2}-i \nu\right)\Gamma\left(\frac{d-1}{2} +i\lambda\right)\Gamma\left(\frac{d-1}{2} -i\lambda\right) \sinh (\pi \kappa) \  h_d(\n,\nu,\lambda)}{2(2\pi)^{1+{d\over 2}}} .
  \label{mfuu}\end{eqnarray}

The evaluation of $ h_d(\lambda,\nu,\kappa)$ is  very far from obvious. In the particular case where the two masses are equal,  $ h_d(\lambda,\lambda,\kappa)$ may be evaluated by Mellin transform techniques, used for the first time in the de Sitter context in \cite{AA,BB}.  
The same idea of using Mellin techniques was used a few years later to compute the  K\"all\'en-Lehmann weight in the case of two equal masses \cite{CC} in AdS QFT\footnote{ At the very same time, however, a general K\"all\'en-Lehmann formula for AdS fields with two different masses was for  the first time published and available \cite{bemads}. But many subsequent authors seem to have ignored it.}. 

The general case of two independent masses cannot be dealt with by Mellin transformation techniques and  something more similar to the Fourier transform of flat space is needed. It is precisely at this point that the plane wave representation  (\ref{uup}) makes a substantial difference.
 
A specially important Fourier-type
representation  is obtained by evaluating  (\ref{uup}) at the purely imaginary
events in the tubes  \cite{bemtri}: $z=-iy\in {\cal T}^-$ and $z = + iy'\in {\cal T}^+$; $y$
and $y'$ can be visualized as points belonging to a  Lobachevsky
space, modeled as the upper sheet of a two-sheeted hyperboloid:
\begin{equation}
\manifold  = \{y\in{\mathbb M}_{1,d}:\ y^2 = y\cdot y = R^2,\;\; y^{0}>
0\}. \label{lbd}
\end{equation}
It follows that 
\begin{eqnarray} w_\nu(-iy,iy') =
{\Gamma({d-1\over 2}+i\nu)\Gamma({d-1\over 2}-i\nu)  \over 2^{
d+1} \pi^d} \int_{\gamma}\left({y\cdot \xi}\right)^{-{d-1\over
2} + i\nu} \left({\xi\cdot y'}\right)^{-{d-1\over 2} -
i\nu}d\mu_\gamma(\xi)  \label{f.14}\cr = \frac{
\Gamma\left(\frac{d-1}{2} +i \nu\right)\Gamma\left(
\frac{d-1}{2} -i \nu\right)}{2(2\pi)^{\frac{d}{2}}}
\left(\left({y\cdot
y'}\right)^2-1\right)^{-\frac{d-2}{4}}\,
P^{-\frac{d-2}{2}}_{-\frac{1}{2} + i\nu}\left( {y\cdot
y'}\right).
\end{eqnarray}
By choosing in particular $\gamma =\gamma_0$ and $y' =
(1,0,\ldots,0)$ so that ${y\cdot y'}= y^0 = u\geq 1$, we then get the
following integral representation:
\begin{eqnarray}
\left(u^2-1\right)^{-\frac{d-2}{4}}\, P^{-\frac{d-2}{2}
}_{-\frac{1}{2} + i\nu}\left(u\right)\ = \frac{1 } {
(2\pi)^\frac d2 } \int_{\gamma_0} \left({y \cdot \xi
}\right)^{-{d-1\over 2} - i\nu} d\mu_{\gamma_0} .
\label{formulaP}
\end{eqnarray} 

This formula is of crucial importance
for computing
$ h_d(\n,\nu,\lambda)$: it allows to rewrite
the one-dimensional integral (\ref{hdn}) as 
the following multiple integral over the manifold $\manifold  \times \sd \times
\sd \times \sd$:
\begin{eqnarray}
&& h_d(\lambda,\nu, \n)= \cr &&
\frac{(2\pi)^{-\frac {3d}2}  } {\omega_{d-1}}
\int_{\gamma_0} \int_{\gamma_0} \int_{\gamma_0}
\int_{\manifold} \left({y \cdot
\xi_1}\right)^{-{d-1\over 2} - i\n} \left({y \cdot
\xi_2}\right)^{-{d-1\over 2} - i\nu} \left({y \cdot
\xi_3}\right)^{-{d-1\over 2} - i\lambda}dy \,
d\mu_{\gamma_0} d\mu_{\gamma_0} d\mu_{\gamma_0} \label{res}  \cr  &&\end{eqnarray}
where $dy$ is the Lorentz invariant measure on $H_d$.
The above integral may be computed and the  final result is  \cite{bemtri}
\begin{eqnarray}
 \rho({\lambda},\nu,\kappa)= \frac{1}{2^{d}{\pi^{{d-1\over 2}}}\kappa
\Gamma\left(\frac{d-1}{2}\right)} \frac{ \ \prod_{\epsilon,\epsilon',\epsilon''=\pm 1}
\Gamma\left(\frac{d-1}{4}
+\frac{i\epsilon\lambda+i\epsilon'\nu +i\epsilon'' \kappa}{2}\right)}
{ \prod_{ \epsilon=\pm 1}
\Gamma
   \left(\frac{i \epsilon \kappa }{2}\right)  \Gamma \left(\frac{1}{2}+\frac{i \epsilon\kappa }{2}\right) \Gamma \left(\frac{d-1}{4}+\frac{i \epsilon \kappa }{2}\right)
   \Gamma \left(\frac{d+1}{4}+\frac{i \epsilon \kappa }{2}\right)}  .
    \cr && 
\label{mf}\end{eqnarray}
Application of this formula and of its AdS twin  to loop calculations are discussed  in \cite{11,12}

\newpage
\section{Anti de Sitter}
The anti de Sitter spacetime 
may be also  visualized as a hypersurface embedded in an ambient flat space which  is $ \bR^{d+1}$  with two timelike directions and metric mostly minus as follows:
\begin{align}
& AdS_d = \{x \in \bR^{d+1}\ :\ x^2= x\cdot x = R^2\}, \label{s.1} \\ 
&  x\cdot y = x^0 y^0-x^1 y^1-\ldots - x^{d-1}y^{d-1} +x^dy^d.
\label{metricads}
\end{align}
The AdS  manifold has a boundary at spacelike infinity and therefore is not globally hyperbolic. This feature gives to  AdS QFT a little extra complication w.r.t. the standard globally hyperbolic case.

We will have to consider also the  complexification of the AdS manifold, which is defined as before  by an embedding:
\beq
AdS_d^{(c)} = \{z =x+iy \in \bC^{d+1}\ :\ z^2 = R^2\}.
\label{complexmanifold}\endq
While  $AdS_d^{(c)}$ is simply connected the real manifold  $AdS_d$ is not and  admits a nontrivial 
universal {covering space}  
$\wt {AdS}_d$. Here  we focus mainly on the uncovered manifold $AdS_d$.

The symmetry group of the anti de Sitter spacetime is is the pseudo-orthogonal group of the ambient space $SO(2,d-1)$. This group may also be regarded as the the conformal group of transformations of the boundary, represented as the null cone of the ambient space
\begin{align}
 C_d = \{\xi \in \bR^{d+1}\ :\ \xi^2= \xi\cdot \xi = 0\}.
\label{nconed}
\end{align}
This simple geometrical fact lies at basis of the AdS/CFT correspondence. The null cone of the ambient space plays also the role of 
giving a causal order to the AdS spacetime  which is however only {\em local}, due to the existence of closed timelike curves;
two events are spacelike separated if
\begin{equation}
(x_1-x_2)^2= 2 -2x_1\cdot x_2 <0. \label{causalads}
\end{equation}
The covering manifold is globally causal but remains non-globally hyperbolic.
\begin{figure}[h]
\includegraphics[width=10 cm]{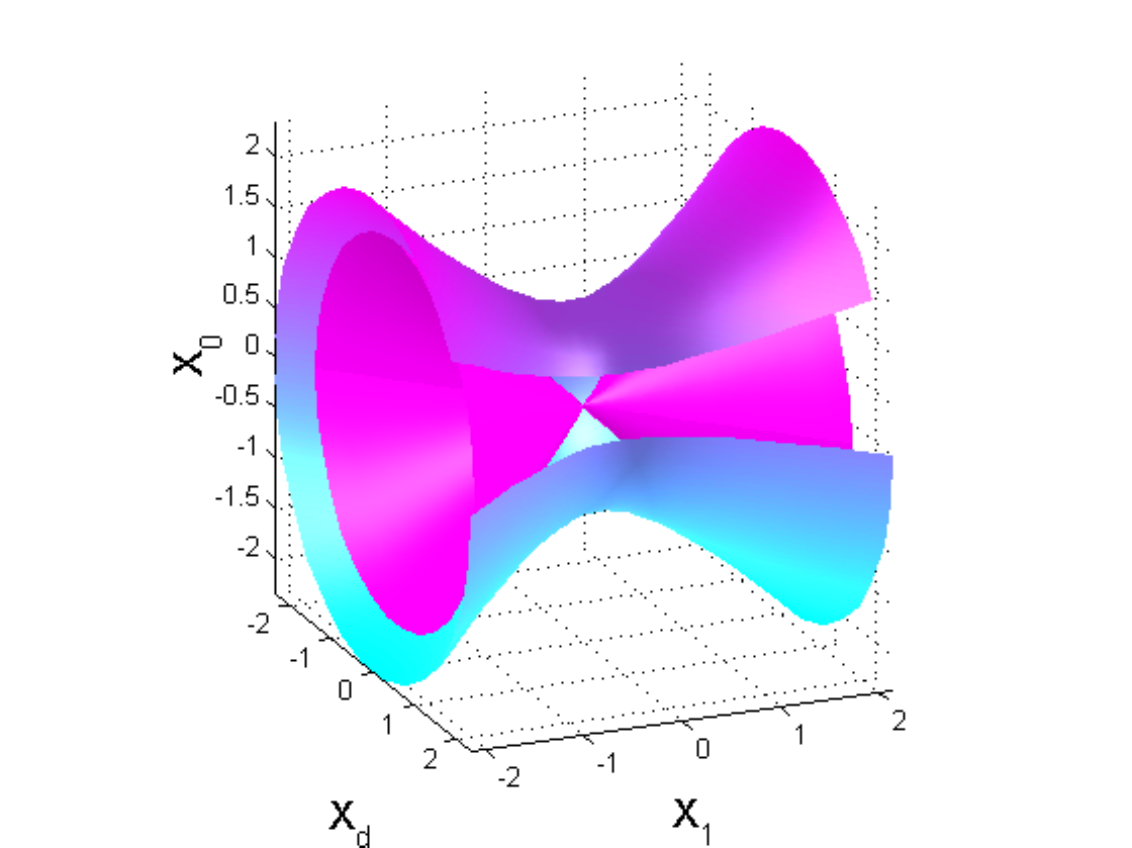} 
\caption{The AdS manifold and the null cone of  the ambient space which models its boundary at spacelike infinity.\label{fig2}}
\end{figure}   
\begin{figure}[h]
\includegraphics[width=12 cm]{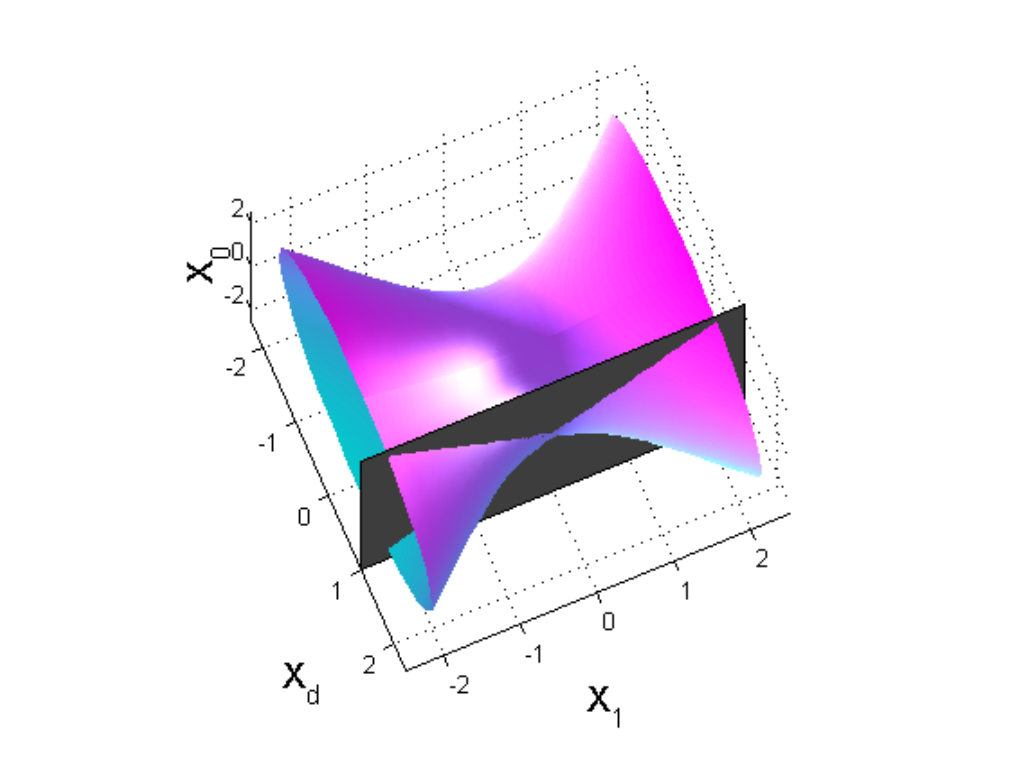} 
\caption{The null cone of  the ambient space induce a causal order on the AdS manifold which is only local. 
\label{fig3}}
\end{figure}   
It is possible to identify in the complex manifold ${AdS}_d^{(c)}$ an analog of 
the Euclidean subspace of the complex Minkowski
spacetime: it is the real submanifold 
$\Lobad$
of ${AdS}_d^{(c)}$ defined by 
\beq
\Lobad = \{z \in \bC^{d+1}\ :\ z\cdot z = R^2, \ \  z(y) =  (y^0,\ldots, y^{d-1},i y^d ),\ y^\mu\in {\bf R }, \ \ y^0 >0\}.
\label{s.12l}\endq
This is indeed a  the same Lobachevsky space we met before in (\ref{lbd})  at the end of the de Sitter tubes, but it has of course a different interpretation in the AdS context, and, more importantly AdS correlation functions have singularities at coincident Euclidean points of $\Lobad$ while dS correlation functions do not.

\subsection{The analytic structure of two-point functions}
The status of  AdS QFT  is more similar to that of Minkowski space and it is possible to formulate a true spectral condition. 
This question has been studied in full generality in \cite{bem2}. A simplified account  may be found in \cite{bbgm}.
The point is that  the parameter of the  group of  rotations in the $(0,d)-$plane
may interpreted as a global time variable: the AdS spectral condition thus is formulated by requiring  that the corresponding generator $M_{0d}$
be represented in the Hilbert space of the theory  by a self-adjoint  operator whose spectrum is positive.
As in flat space, this requirement is equivalent to precise  analyticity properties of the 
of the $n$-point functions  \cite{bem2}. 

In particular, there are two distinguished complex domains of
$AdS^{(c)}_{d}$, invariant under real AdS transformations  \cite{bem2,bbgm} defined as follows
\begin{eqnarray} 
{\cal Z}_+ =\{ z  = x+iy  \in AdS^{(c)}_{d}; \, y^2>0,\, \epsilon(z) = +1\},\\
{\cal Z}_-  =\{ z = x+iy  \in AdS^{(c)}_{d}; \, y^2>0,\, \epsilon(z) = -1\},
\label{Tubes}
\end{eqnarray}
where
\begin{equation}
\epsilon(z) =  \mbox{sign}  (y^0 x^{d}- x^0 y^{d}).
\end{equation}
The tubes ${\cal Z}_+$ and ${\cal Z}_-$ have a definite chirality and wrap the real  AdS manifold in opposite directions.
{ The spaces  ${\cal Z}_+$, ${\cal Z}_-$ and $AdS_d$ have the
same homotopy type.}
Their universal coverings are denoted $\wt{\cal Z}_+$  and $\wt{\cal Z}_-$.

The AdS spectral condition implies that   a general
two-point function  satisfies the following \cite{bem2}
 \vskip 10 pt
\noindent 
{ \bf Normal analyticity property:}
{\em $ \W\left( x_1,x_2\right)$
is the boundary value of a  function $ W\left(
z_1, z_2\right) $ holomorphic in the  domain
${\cal Z}_-\times {\cal Z}_+$} 
\begin{equation}
\W(x_1,x_2) = \langle \Psi_0, \widehat\phi(x_1)\widehat  \phi(x_2)\Psi_0 \rangle =\underset{\underset{  {\cal Z}_+ \ni \, z_2 \to x_2 }{\scriptscriptstyle {\cal Z}_- \ni \, z_1\to x_1 }}{ b.v.} {{ W}}(z_1,z_2).
\end{equation} 
 \vskip 10 pt
 
\begin{figure}{h}
\includegraphics[width=13cm]{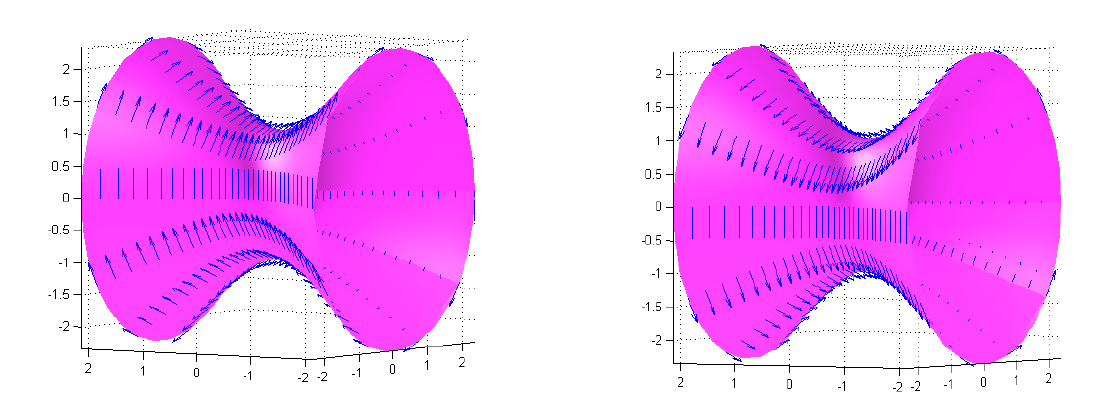} 
\caption{Sections of the backward and forward tubes in the complex AdS manifold. 
The tubes wrap the real manifold in opposite directions.
The arrows are the imaginary parts $y$ of the vectors 
$z=x+i y$ parts attached at the end of the real parts $x$ which varies on the hyperboloid whose radius is $R^2+y^2$. Here is represented the section at a fixed positive value of $y^2$. Recall that $x\cdot y =0$. \label{fig4}}
\end{figure}

\noindent AdS invariance and normal analyticity  imply the following 
 \begin{theorem} [Maximal analyticity property]
\begin{enumerate}\item The two-point function $W(z_1,z_2)$  depends only on the AdS-invariant  variable $\zeta = z_1\cdot z_2$.
\item $W(z_1,z_2)$ can be continued to a function $ {\frak W}(z_1,z_2)$ analytic in the cut-domain
\begin{equation}\Delta_1 = \{{\bf C} \setminus[-1,1]\}.\ \ \ 
\end{equation}
 
\item $ {\frak W}(z_1,z_2)$ is invariant in under the action of the complex de Sitter group. 

\item The permuted two-point function
is the boundary value of  ${\frak W}(z_1,z_2)$
from the opposite tube $
{ \cal Z}_{21} = {\cal Z}_+\times {\cal Z}_-$.
\item  The cut-domain $\Delta_1$  contains the all the non-coinciding Euclidean points 
\begin{equation}
\dot{\cal E}= \{z_1, z_2 \in \Delta, \ \ z_1 \in S_d , \  z_2 \in H_d , \ \ z_1\not=z_2 \}.
\end{equation}
The Schwinger function $S$  is the restriction of ${\frak W}(z_1,z_2)$ to the non-coincident Euclidean points
 $\dot{\cal E}$.  $S$ is  analytic in  $\dot{\cal E}$ and can be extended as a distribution to the whole Euclidean space ${\cal E}$ including the coinciding points.
\end{enumerate} 
\end{theorem}
As regards the global hyperbolicity issue, the  maximal analytic structure  
completely  determines the   two-point functions for Klein-Gordon fields and, as a consequence, selects the boundary behaviour of the modes.

\subsection{Klein-Gordon fields and plane waves}
Klein-Gordon fields display
the simplest example of the previous analytic structure. For a given mass $m$ the two-point function  ${\cal W}(x_1,x_2)$ must satisfy the equation 
\begin{equation}
(\square_{x_i} + m^2) {\cal W}(x_1,x_2) = 0, \ \ i=1,2,\label{kg}
\end{equation}
w.r.t. both variables, 
where $\square_{x_i}$ is the Laplace-Beltrami operator relative to the AdS metric. 
The two-point functions are labelled by the (complex) dimension $d$ and a (complex) parameter $\nu$
as follows
\begin{align}
&W^d_\nu(z_1,z_2)=w^d_{\nu}(\zeta) =
{1 \over (2\pi)^{d\over 2}} (\zeta^2-1)^{-\frac{d-2}4}
e^{-i\pi{d-2\over 2}} Q_{-{1\over 2}+\nu}^{d-2\over 2}(\zeta)=
\label{s.9}\\
&=\frac{ \Gamma
   \left(\frac{d-1}{2}+\nu \right)  \,
  }{ 2\pi ^{\frac{d-1}{2}}  (2\zeta)^{ \frac{d-1}{2}+\nu }\Gamma (\nu +1)}  \ {}_2F_1\left(\frac{d-1}{4}+\frac{\nu }{2},\frac{d+1}{4}+\frac{\nu }{2};\nu
   +1;\frac{1}{\zeta^2}\right)
\end{align}
where the various parameters are related as follows:
\begin{equation}
     m^2 = \nu^2 -\frac{(d-1)^2}4.
\end{equation}
There  are two possible cases : 
\vskip 10pt

1. for $\nu > 1$ the spectrum condition uniquely select one field theory for each given value of mass parameter $\nu$;
\vskip 10pt

2. for $|\nu| < 1$ there are two acceptable theories for each given
mass. 

The difference
between the two theories is in their
large distance behavior; more precisely, in view of  \cite[Eq. (3.3.1.4)]{bateman} one has that 
\begin{equation}
w^d_{-\nu} (\zeta)  = w^d_{\nu} (\zeta)  + \frac{\sin
\pi\nu \, \Gamma\left(\frac {d-1} 2-\nu\right) \Gamma
\left(\frac {d-1} 2 +\nu\right) }{(2\pi)^{\frac{d}{2}}} \, (\zeta^2-1)^{-\frac {d-2}4}P^{-\frac
{d-2}2}_{-\frac 1 2 -\nu}(\zeta) . \label{pop}
\end{equation}
The last term in this relation is {\em regular on the cut}
$\zeta \in [-1,1]$ and therefore does not contribute to the
commutator. By consequence the two theories represent the same
algebra of local observables at short distances. But since the
second term at the rhs grows the faster the larger is
$|\nu|$ (see \cite[Eqs. (3.9.2))]{bateman} the two theories have 
drastically different  long range behaviors.
\vskip 10 pt

Let us now proceed to the harmonic analysis in plane waves also for the AdS correlation functions. 
Here, to keep the discussion as simple as possible, we limit ourselves to the two-dimensional uncovered Anti de Sitter spacetime $AdS_2$ \cite{bmf}. 
A full analysis will be presented elsewhere.

As for the de Sitter case, also the wave solutions of the anti de Sitter Klein-Gordon equation
may be 
parameterized by the choice of a null vector $\xi\in C_2$  and a complex number $\lambda$  as follows:
\begin{equation}
\psi_\lambda(z,\xi) = \left({ z\cdot \xi}\right)^{\lambda}  = e^{\lambda \log \left({ z\cdot \xi}\right) }   \label{wavesads}.
\end{equation}
Since we are consideiring the uncovered manifold, the parameter $\lambda$ must  be an integer: 
\begin{equation}
\lambda= \ell.
\end{equation}
Now we observe that, while  
Eq. (\ref{geod}) maintains its validity also in the present context, for real $\xi$, $\eta$ belonging to the null cone $C_2$ it describes spacelike geodesics. 
Timelike geodesics would correspond to vectors $\xi$ belonging to  the complex cone
\begin{align}
 C_2^{(c)} = \{\xi \in \bC^{3}\ :\ \xi^2= \xi\cdot \xi = 0\}.
\label{ncone}
\end{align}
This suggests that the harmonic analysis of the AdS correlation function should also be made in terms of waves parametrized by null complex vectors $\xi$.

The complex cone $C^{(c)}_2 $ admits the following partition:
$C_2^{(c)} = C_2 \cup C_{2+} \cup C_{2-}$ where
\begin{align}
& C_{2+} = \{ \xi \in C^{(c)};\ \epsilon(\xi) = +\} ,\\
& C_{2-} = \{ \xi \in C^{(c)};\ \epsilon(\xi) = -\};
\end{align}
as before
\begin{equation}
\epsilon(z) =  \mbox{sign} [ (\Im \xi^0) (\Re \xi^2)-(\Re \xi^2)( \Im \xi^2 )].
\end{equation}
Bases  for the cones $C_{2+} $ and $C_{2-}$ are 
\begin{align}
&(\gamma_0^{(c)})_{+} = \{\xi =\xi(\Phi)= (\sin \Phi,1,  \cos \Phi);\
\Phi = \phi + i \eta,\  \eta >0 \},\\ 
&(\gamma_0^{(c)})_{-} = \{\xi =\xi(\Phi)= (\sin \Phi, 1,\cos \Phi);\
\Phi = \phi + i \eta,\  \eta <0 \}.
\end{align}
Let us now consider the integral 
\begin{equation}
\int_ {\gamma(z_1)}
[z_1\cdot \xi]^{\ell} \
[\xi\cdot z_2]^{-\ell -1}
{\rm d}\mu_{ \gamma^{(c)}} (\xi), \ \ \ z_1 \ in {\cal Z}_-, \ \ \ z_2 \ in {\cal Z}_+.
\label{legendre2}
\end{equation}
For  for each $z_1$ in
${\mathcal Z}_-$, 
$\gamma(z_1)$ is a relative cycle in $H^1(C_2^{(c)}, \{\xi;\ [z\cdot \xi] =0 \})$ with support contained in
$C_{2-}$ and end-points  belonging respectively 
to the two linear generatrices of the cone $C_2^{(c)}$ defined by  $[z_1\cdot \xi] =0$.
Being the integral of a closed differential form, (\ref{legendre2})  not depend on the choice of 
$\gamma(z_1)$ inside its homology class.

There is no loss of generality in  defining the integration cycle $\gamma(z_1)$ only for points of the form 
\begin{equation}
z_1= z_v = (i \sinh v,0,\cosh v), \ \ v<0.
\end{equation}
We choose  the path 
\begin{equation}
\phi \rightarrow\xi (\phi+i v)=  (\sin (\phi +iv),1,  \cos(\phi +iv)), \ \ -\frac{\pi}{2}<\phi<\frac{\pi}{2}.
\end{equation}
The support of $\gamma(z_v) $  does belong  to $C_{2-}$ and $[z_v\cdot \xi(\phi+i v)] = \cos(\phi)$ vanishes, a required,  at the boundaries of the cycle.

Since $z_2 \in {\cal Z}_+$ the  
factor $[\xi\cdot z_2]^{-\ell -1}$ never becomes singular on the integration cycle $\gamma(z_v)$. This may me seen by explicitly giving coordinates to ${\cal Z}_+$ \cite{bmf}. 
This suffices to show 
the AdS invariance  of the integral
(\ref{legendre2})
which is therefore a function of the invariant  variable $\zeta= z_1\cdot z_2$, 
holomorphic in the cut domain $\Delta_1$

To actually compute (\ref{legendre2})
we choose the second point at the origin $x_0=(0,0,1)$ 
so that $z_1\cdot z_2 =z_v\cdot x_0= \cosh v$.
With a few self-explanatory changes of variables we get  \cite[Eq. 2, p. 155]{bateman}
\begin{align}
&\int_ {\gamma(z_v)}
[z_v\cdot \xi]^{\ell} \
[\xi\cdot x_0]^{-\ell -1}
{\rm d}\mu_{ \gamma} (\xi) 
= \cr&=
\int_{-\frac{\pi}{2} }^{\frac{\pi}{2}}
cos^{\ell} \phi\  \cos^{-(\ell +1)} (\phi + i v) \ \ {\rm d}\phi
= 
\int_{-\infty}^{\infty} (\cosh v -i t \sinh v)^{-(\ell +1)} \frac {{\rm d}t}{(1+t^2)^{\frac{1}{2}}} \cr 
&= 2\int_{0}^{\infty} (\cosh v +\cosh u \sinh v)^{-(\ell +1)} du = 2 Q_l(\cosh v).
\label{leg2}
\end{align}
Therefore we may write the following plane-wave expansion of the two-point function 
\begin{align}
&W^2_{l+\frac 1 2}(z_1,z_2)= {1 \over \pi} \int_ {\gamma(z_v)}
[z_v\cdot \xi]^{\ell} \
[\xi\cdot x_0]^{-\ell -1}
{\rm d}\mu_{ \gamma} (\xi) .
\end{align}

\section*{Acknowledgments}{I would like to thank the Department of Theoretical Physics of CERN and its director Gian F. Giudice for warm hospitality and support while  writing this paper.}





\end{document}